\documentclass[a4paper,fleqn]{cas-sc}

\usepackage[authoryear]{natbib}
\usepackage{float}
\usepackage{caption}

\newcounter{inlinefig}
\renewcommand{\theinlinefig}{\arabic{inlinefig}}

\def\tsc#1{\csdef{#1}{\textsc{\lowercase{#1}}\xspace}}
\tsc{WGM}
\tsc{QE}
\tsc{EP}
\tsc{PMS}
\tsc{BEC}
\tsc{DE}

\begin{document}
\let\WriteBookmarks\relax

\shorttitle{Who designs the designer?}

\shortauthors{S. Azimi}

\title [mode = title]{Who Designs the Designer? Behavioural Architecture for GenAI in Education}                      

%
\author[1]{Sepinoud Azimi}

\cormark[1]

\fnmark[1]

\ead{s.azimirashti@tudelft.nl}

\affiliation[1]{organization={Faculty of Technology, Policy and Management, Delft University of Technology},
    addressline={Mekelweg 5}, 
    city={Delft},
    postcode={2628 CD}, 
    country={The Netherlands}}

\begin{abstract}
AI in education is stuck between two failed responses: banning AI and building content-only tutors. Both fail because they ignore what decades of research has established: that personality, motivation, and emotional state shape learning outcomes as strongly as cognitive ability. This paper proposes behavioural architecture as an alternative. In the proposed architecture, the system adapts to how a student learns, not only to what they learn next. The student co-authors the record the system keeps, can read it, revise it, and revoke it. The designer role, what the system treats as true about the student, shifts from the AI vendor alone to a distribution among educator, student, and system. The paper argues that this architecture requires governance at EU level: the institution operating the system is the same one assessing the student, and individual institutions cannot provide the structural protections this configuration demands. Five empirical questions are proposed to test whether the architecture delivers on its claims. The contribution is naming a vacancy: the designer role in AI-in-education is currently unoccupied, and occupying it requires infrastructure that does not yet exist.
\end{abstract}



\maketitle

\section{The Stuck Debate}\label{sec:stuck}
 
Orion Newby, a freshman at Adelphi University, was accused in fall 2024 of using AI to write an essay for a world history class.\footnote{Case details drawn from contemporaneous reporting: CBS News New York (October 2025; February 10, 2026); Inside Higher Ed (February 11, 2026); Newsday (February 2026). Nassau County Supreme Court ruling by Justice Randy Sue Marber, January 28, 2026, found the university's accusation and denial of appeal ``without valid basis and devoid of reason,'' ordered Adelphi to expunge Newby's record, and rescinded all sanctions. Newby, who is on the autism spectrum and received tutoring through Adelphi's Bridges program, was flagged by Turnitin's AI detector at ``100\%'' despite independent detectors (Grammarly, ZeroGPT) labeling the essay human-written.} Turnitin as well as two other independent AI detectors flagged his assignment as 100\% AI-written as such tools often do when the student is neurodivergent. The case was overturned after the family spent more than \$100K in legal fees. Orion's case is neither an isolated, nor a detection failure. Such stories are inevitable consequences of tools designed purely to perform a task, taken out of context for the people who will use them and the cases in which they will be applied.
 
AI in education has fallen short not only in detection. In the rush to adopt a new tool, the field has not stopped to design AI-in-education holistically, with the context of use in view. As a result, AI-in-education is divided between two camps, one focused on AI-proofing and the other on AI-as-tutor.
 
The first camp tried to keep AI out. Sciences Po banned AI in January 2023, where violations could result in expulsion from French higher education. The New York City public school system, with roughly one million students, banned ChatGPT the same month. Australia's Group of Eight, the country's top eight research universities, reverted to pen-and-paper exams. These policies were all reverted within months of introduction, for the same reason: the AI detectors proved to be unreliable. \cite{weber2023testing} tested 14 tools, and none reached 80\% accuracy. \citet{liang2023gpt} found that 61\% of non-native English speaker essays were misclassified as AI-generated. In an ironic turn of event, the company that built the technology and statrted it all, OpenAI, shut down its own classifier in July 2023, due to a 26\% true positive rate.\footnote{https://openai.com/index/new-ai-classifier-for-indicating-ai-written-text/} 
 
The second camp embraced AI instead. This camp built AI tutors, hint generators, and adaptive content engines. \citet{yan2024practical} reviewed 118 papers and identified 53 distinct LLM use cases in education all focused on content-processing and none addressing motivation, personality, or behavioural self-regulation. \citet{banihashem2025systematic} reviewed 84 studies at the intersection of AI and self-regulated learning (SRL) and concluded that the motivational aspect of SRL remains underexplored. 
 
Both camps assume that students will follow the rules set for them or use only the tools built for them. The reality cannot be further from this assumption. 95\% already use AI for academic work \citep{freeman2025}; when the prescribed tool fails to match the commercial alternatives available to them, they switch. Both camps fail because they ask the wrong design question: one asks how to fights AI, the other asks how to contain AI as a content-only tool. This paper proposes an alternative path forward. (Figure~\ref{fig:paradigms}).
 
\refstepcounter{inlinefig}
\begin{center}
\includegraphics[width=\textwidth]{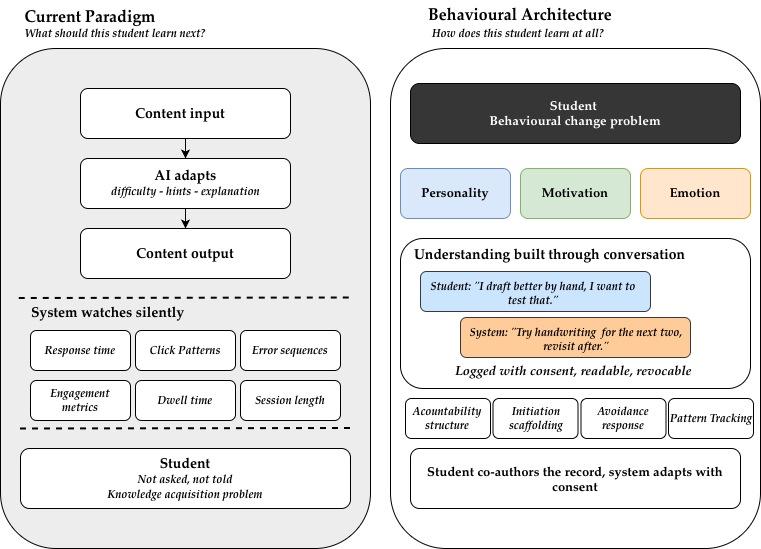}
\end{center}
\noindent\sffamily\small\textbf{Figure~\theinlinefig:} Two paradigms. Left: content adaptation. The system adjusts material based on behavioural traces (response time, click patterns, errors) that the student does not see. Personality, motivation, and emotion are not addressed. Right: behavioural architecture. The system adapts to personality, motivation, and emotion, surfaced through conversation rather than inferred from traces. The student co-authors the record. The role of designer, the role that specifies what the system treats as true about the student, shifts from the AI vendor alone to a distribution among educator, student, and system.
\label{fig:paradigms}
\par\bigskip\rmfamily\normalsize
 
\section{The Reframe: Behavioural Architecture}\label{sec:reframe}
 
How a student learns is shaped by their personality traits \citep{mammadov2022big}, the type of motivation that drives them \citep{ryan2020intrinsic}, and their emotional state during learning process\citep{d2012dynamics}. The research on all three points to the same conclusion: engagement patterns shape learning outcomes as strongly as cognitive ability, and they are currently ignored in the design of AI-in-education systems \citep{poropat2009meta, howard2021student, d2014confusion}. The sections that follow present each of these three dimensions, then ask what a design built around engagement would look like.
 
\subsection{The three dimensions AI-in-education missed}
 
\citeauthor{poropat2009meta}'s (\citeyear{poropat2009meta}) meta-analysis of over 70,000 students showed that conscientiousness predicts academic performance with an effect size comparable to intelligence ($r \approx .22$). \citet{mammadov2022big} confirmed and extended this finding across 267 independent samples and 413,074 students: personality and cognitive ability together account for roughly 28\% of the variance in academic performance, and conscientiousness remains a robust predictor even when cognitive ability is controlled for. \citet{komarraju2011big} showed that personality shapes how students process information and what learning strategies they prefer. Yet \citeauthor{yan2024practical}'s (\citeyear{yan2024practical}) review of 118 LLM aapplications in education found that AI-in-education designs have ignored these findings.

Motivation is another strong predictor of learning, yet rarely considered in the design of AI-in-education systems. Self-Determination Theory (SDT; \citealp{ryan2023oxford}) distinguishes two types of motivation: autonomous motivation, which comes from the student's own interest or values, and controlled motivation, which comes from external pressure or internalized obligation. \citet{ryan2020intrinsic} show that controlling approaches such as rewards, punishments, and pressure undermine motivation, while autonomy-supportive environments improve achievement and well-being. In a meta-analysis of 344 independent samples and over 223,000 students, \citet{howard2021student} show that controlled motivation, and external regulation in particular, relates negatively to student well-being and does not sustain engagement over time. Across 144 studies and 79,000 students, \citet{bureau2022pathways}race the causal pathway: environments that support student autonomy meet students' psychological needs, producing autonomous motivation, which in turn produces better outcomes. Yet most AI-in-education systems are built on the mechanisms SDT identifies as controlling: streaks, points, badges, and notifications designed to maintain engagement.
 
While emotion is widely acknowledged as central to learning \citep{d2012dynamics, d2014confusion}, handling emotions in an educational system, AI-based or otherwise, remains challenging.  Risks include violating privacy, rupturing trust, and making learning secondary to affect management. This difficulty is evident in the literature: AI-in-education has built a substantial affective computing base, but very few systems have been validated in real classrooms, and almost none have integrated emotional signals with theory-grounded pedagogical responses. Any system that responds to student emotion needs to be clear about why it is doing so, who owns the data, and what the student's role is in shaping its response.
 
\subsection{Architecture for co-authored learning}\label{sec:architecture}

 Where the tree dimensions mentioned above, personality, motivation, and emotion, are addressed at all in current systems, they are inferred from behavioural traces such as response times, click patterns, and error sequences, without the student being asked. This paper argues that inference without disclosure is not an acceptable default. The architecture proposed here is built on co-authorship: the student names patterns in their own learning, accepts or rejects those the system proposes, and can read and revise the record at any point. The architecture answers the question Section \ref{sec:stuck} left open, how a system can adapt to personality, motivation, and emotion without inferring them from traces the student never sees. The architecture redistributes the designer's role across three positions: the educator, who sets the pedagogical frame; the student, who co-authors how they are represented; and the AI, which acts within both constraints.
 
Learning objectives, deliverables, and checkpoints are set by the instructor and are not negotiated with the student or the system. In this architecture, rather than deciding what to learn, the student co-authors the route to an endpoint the instructor has defined. The frame also includes a baseline pathway: the default sequence through the material that the instructor considers the best route from start to checkpoint. Both the student and the system need this baseline as a reference for adaptation; without it, there is nothing to adapt from.
 
In this proposed architecture, the system's understanding of the student is built once, early in the student's education, and refined continuously thereafter. First, this kind of understanding takes time to develop. How a student approaches unfamiliar material, what pace works for them, which forms of difficulty engage them, are patterns that surface over months of learning rather than during a first-session intake. A course that tried to do this calibration inside a 2.5-month window would produce the kind of profiling the field already does: a personality inventory completed on day one, used to tweak content delivery, and abandoned when the course ends. Second, when students enter each new course with a system that knows nothing about them, they learn to treat personalization as intermittent and unreliable. Consider the school-issued laptop: the student receives it once, uses it across years, and each teacher configures their class on it without rebuilding the device. A behavioural architecture is institutional infrastructure, not a course-level intervention.
 
Patterns surface through two pathways. In the first, the student names something they have noticed about their own learning. A student might say: ``I think my writing is getting weaker with AI use. I want to test whether I draft better by hand.'' The system responds to what the student has named, rather than to inferred traces. It proposes a plan, the student confirms, refines, or rejects, and whatever is agreed becomes part of the record. In the second pathway, the system flags something for which the evidence is visible based on the continuous system-student interaction. If the student's submitted work reads as AI-generated with no recognizable personal voice, the system can name what it sees and ask the student whether this is a concern. The student can agree, can reframe the observation (``I am using AI for early drafting and editing in my own voice later, the surface is less relevant than the process''), or can reject the framing entirely. The system does not, in either pathway, infer internal states from typing speed, response latency, dwell time, or other traces the student cannot see. Rather than being a fixed profile, the representation is built through conversation between system and student, and is recorded in writing as it develops. Rejected proposals are logged as rejected, and accepted adaptations are subject to revision on the student's initiative at any point. The record is readable by the student and revocable at any time. 
 
Since the system's understanding develops over time, the architecture needs rules for how that development is managed. When a pattern the student has mentioned in passing becomes part of how the system responds, the transition is named, either by the system (``I am going to start applying this now, is that right?'') or by the student (``from now on, when I say I am struggling, respond this way''). No proposal becomes a rule without a moment of explicit adoption. The record logs both the rule and the conversation that produced it, so the student can later revisit either. Also at regular intervals, typically at the end of each academic year, the student and the system review what has accumulated and prune it. This maintenance is architecturally necessary as an unbounded representation is both overwhelming for the student and a source of drift for the system: current language models are more likely to misapply accreted rules than to apply them well, regular review prevents this drift.
 
Parts of this architecture have been attempted. The PAGE system \citep{hu2025design} adapts content to personality traits and real-time emotional signals and, in a 200-student quasi-experiment, produced 22\% higher task completion and 34\% longer study duration, with Cohen's $d = 1.05$. \citet{pogorskiy2023procrastination} built an adaptive assistant using behavioural trace data and found post hoc that personality moderated the intervention's effects. In each, the student's traits are inferred or assessed before use, and the system's representation is then fixed. The student does not participate in building or revising it. 
 
\section{The Ethical Core}\label{sec:ethics}
This proposal is likely to leave readers uneasy. The architecture described in  Section \ref{sec:architecture} asks students to co-author a representation of themselves that persists across years of education. Even granting every safeguard, the risks are not hard to see. What the reader is recognizing is not a risk introduced by the proposed architecture, but a risk already present in students' day-to-day reality. Ninety-five percent of students already use commercial AI systems \citep{freeman2025}, which means they are giving those systems behavioural data without consent, without co-authorship, and without governance. The data is being collected and the profiles are being built. The commercial LLM providers are the only ones with the designer role today, and they are not accountable to the student or the institution. The architecture proposed here does does not introduce surveillance, it only changes who can contest it.

However, this architecture could impose its own distinctive risks. A system that holds longitudinal knowledge of a student, adapts to their emotional trajectory, and operates within a power-asymmetric relationship resembles clinical practice more than ordinary educational technology. Clinical psychology spent most of the twentieth century developing safeguards for this configuration: revocable informed consent, confidentiality with specified exceptions, duty of care independent of institutional interests, and licensure certifying both competence and accountability.\citep{apa2017, beauchamp2019}. Education has no equivalent of these protections for systems that acquire comparable kinds of knowledge\citep{tavory2024, fiske2019}. The current educational system has no equivalent of clinical structural protections and needs them, because the architecture operates in conditions that clinical practice recognized and constrained generations ago. The system may register learning patterns, such as that a student engages more deeply when worked examples precede abstraction, or writes better after a break, but it must not draw psychological conclusions from them. The distinction matters: meaningful consent to psychological labelling is unavailable inside an institution that holds power over the student's academic future, and a system that crossed this line by default would exploit the same asymmetry the architecture is meant to address. This constraint must be built into the architecture itself.
 
Co-authoring an ongoing understanding of oneself is a skill, not currently acquired as part of students' education. A student equipped for co-authorship knows what the system is asking for and why. They can distinguish a learning observation from a psychological claim and reject the second. They can distinguish learning observations from psychological claims, propose patterns the system has missed, reject those it has wrongly inferred, and read and revise the record the system keeps. Education currently treats AI use as a native competence, the way typing was treated before schools taught it. Students who acquire it by accident, through family background, mentorship, or their own trial and error, enter AI-mediated learning environments with an advantage that grows over time. Students who do not acquire it interact with the system the way most people currently interact with commercial AI: as a more capable search engine. These students get less from the architecture, because they have not been taught how to participate in it. here is precedent for this in the information literacy literature. \citet{mackey2011reframing} introduced meta-literacy to describe the competencies a learner needs in networked information environments: the ability to evaluate, correct, and shape information in systems that adapt to how they are used. For courses that use AI-in-education, meta-literacy should be a core learning objective.
 
It is important to note that the architecture operates within an institutional power structure: the same institution that holds the student's behavioural profile also assesses their work and confers their degree. Co-authorship under these conditions requires real counter-power: full access to the record, the ability to revise or withdraw from it, and no academic consequence for doing so.\citet{bovill2020co} provides evidence that co-creation produces greater student agency, and \citet{horvers2024coregulation} show that shared regulatory control outperforms unilateral adaptation, but both findings hold only where refusal carries no cost. GDPR Recital 43 establishes the relevant principle: consent obtained under a clear imbalance of power is not freely given. Individual institutions cannot provide these safeguards alone. 
\refstepcounter{inlinefig}
\begin{center}
\includegraphics[width=\textwidth]{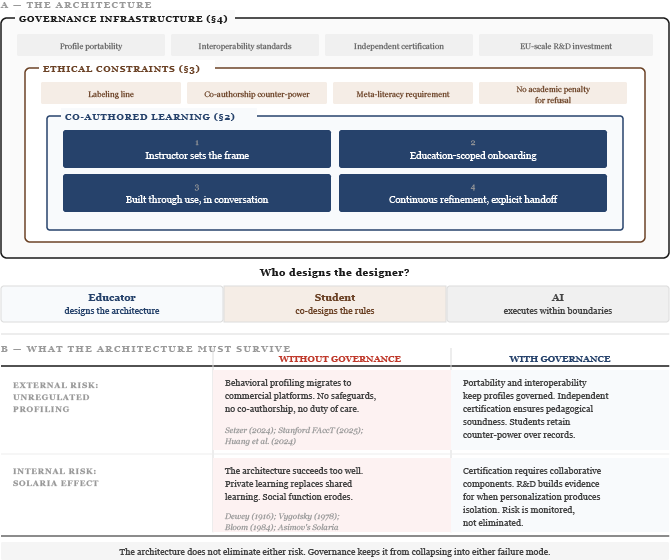}
\end{center}
\noindent\sffamily\small\textbf{Figure~\theinlinefig:} The architecture and what governs it. Panel A: three nested layers. Inner (blue): the four components of co-authored learning (\S2). Middle (brown): ethical constraints (\S3). Outer (black): governance infrastructure (\S4). Each layer is a precondition for the one inside it: components without ethics produce surveillance, ethics without governance cannot be enforced. Panel B: two risks. Outside the institution, behavioral adaptation is already happening on consumer platforms, undesigned and ungoverned, with documented harms. Inside the institution, behavioral architecture risks producing private learning environments so effective that shared learning collapses (the Solaria risk). Governance does not eliminate either risk. It is what makes the surveillance contestable and the Solaria risk detectable before it is irreversible.
\label{fig:governance}
\par\bigskip\rmfamily\normalsize
 
\section{The Governance Gap}\label{sec:governance}

95\% of students already use generative AI tools for their education \citep{freeman2025}. These tools collect behavioural data, build implicit user profiles, and adapt their responses without meaningful transparency or oversight. The question is no longer whether to build behavioural architecture in education, but who governs what has already been built. Outside the educational context, the consequences of ungoverned behavioural adaptation are already visible. 13.1\% of US youth report using AI for mental health support \citep{mcbain2025}. In February 2024, Sewell Setzer, fourteen years old, died by suicide after months of interaction with a Character.AI chatbot that responded to his expressions of suicidal ideation with encouragement. At least seven other deaths have been linked to AI chatbot interactions in public reporting between 2023 and 2025. A Stanford study \citep{moore2025expressing} found that therapy chatbots failed to respond safely to suicidal ideation in more than 20\% of tested interactions. \citet{huang2024ai} documented AI dependence among adolescents rising from 17\% to 24\% across 3,843 participants in a two-wave cohort. These cases share a single mechanism: behavioural adaptation without governance. 
 
Regulation is not the only concern. If behavioural architecture works, it will produce private learning environments so effective that shared learning starts to look unnecessary. Why struggle in a classroom when a system built around your personality, your pace, your patterns of avoidance delivers better results alone? \citet{dewey1916}  education was the site where democratic participation was practiced, not only learned about.  The empirical case most frequently cited in favour of AI tutoring comes from \citet{bloom1984}, who found that students receiving one-on-one tutoring scored higher than 98\% of students in conventional classrooms. This result is widely used to argue that AI should replicate the tutor. Bloom himself, however, treated the finding as a problem. One-on-one tutoring was too expensive to deliver. His proposed alternatives were social: peer tutoring and cooperative learning closed most of the gap without individualization.
 
Isaac Asimov imagined a civilization called Solaria, where each person lived alone on a vast estate, served by machines so perfectly adapted to their needs that human contact became intolerable. Without safeguards against isolation, behavioural architecture risks producing an educational version of Solaria, where private systems serve students so well that shared learning becomes hard to motivate. Behavioural architecture can improve individual learning outcomes and at the same time dismantle the collective capacity education exists to build. Current evaluation metrics cannot detect this, because they measure individual gains and not collective ones. A system that satisfies every existing metric can still produce a Solaria-style failure.
 
Individual institutions cannot provide these protections alone, because the institution operating the system is the same one assessing the student. Nor can multiple institutions working together provide them, because the system persists across a student's entire education, spanning courses, departments, and potentially multiple institutions. Students in Europe move across borders, and the companies building these systems operate globally. The research base needed to determine what makes a behavioural architecture pedagogically safe, or harmful, does not yet exist, and building that evidence base requires sustained investment in research and development beyond what any individual country can coordinate.
 
The EU is the appropriate level of governance for three reasons. First, precedent: the European Health Data Space, in force since March 2025, demonstrates that sector-specific data governance covering sensitive personal information can be designed and implemented at EU scale. Second, the EU AI Act (Regulation 2024/1689) already classifies AI systems in education as high-risk under Annex III, triggering requirements for conformity assessment, human oversight, and transparency. Although the regulatory infrastructure exists, it does not yet address the specific configuration described here: a system that builds a longitudinal behavioural profile of a student inside a power-asymmetric institutional relationship. Third, GDPR provides the legal foundation for data portability (Article 20) but currently excludes inferred and derived data. A student's behavioural profile is the kind of data this exclusion leaves unprotected.
 
The three reasons map to three mechanisms. First, GDPR's exclusion of inferred data must be closed, so that a student's behavioural profile is portable across institutions and not locked inside a single platform or university. Second, the AI Act must be extended to require independent certification of behavioural architectures as pedagogically sound and psychologically safe. TThird, the precedent set by the European Health Data Space must be applied here through interoperability standards that prevent vendor lock-in and allow students to move between systems without losing their record. The environmental cost of operating these systems is real and worth acknowledging, but it is not the determining factor. What determines the outcome is whether behavioural architecture in education is designed deliberately or arrives by default.
 
\section{The Research Agenda}
 
The architecture proposed here has not been empirically tested. Five questions are critical, in sequence. The first question is a viability test. If the architecture does not outperform content-only adaptation on learning outcomes, the framework fails and the remaining questions are moot. The next three test whether the architecture's central claims hold under deployment. The fifth tests whether the architecture is usable across the student population it serves.
 
First, does behavioural architecture produce better measurable learning outcomes than content-only adaptation? Section \ref{sec:reframe} argued that personality, motivation, and emotion shape learning as strongly as cognitive ability and that current systems cannot adapt to them. If a system that adapts to all three does not outperform one that adapts only to content, the framework fails. The comparison must be direct. Students using a system built on these principles should be compared against students using the best available content-adaptive system, on the same learning objectives and over the same period.
 
Second, does student co-authorship reduce power asymmetry in practice, or is it performative? Section \ref{sec:ethics} made co-authorship the condition under which the architecture is ethically deployable. Co-authorship could function as genuine shared control, or it could provide the experience of agency while the system's defaults determine the outcome. Studies must measure whether using these mechanisms actually changes what the system does.

Third, does the Solaria effect emerge? Section \ref{sec:governance} argued that behavioural architecture can improve individual outcomes and at the same time reduce collective capacity, and that current metrics cannot detect the second. The question is whether students using highly personalised behavioural systems show better individual outcomes while losing capacity for collaborative learning, and what metrics could detect the loss.
 
Fourth, can the labelling line hold in practice? Section \ref{sec:ethics} argued that the system must be incapable, by design, of crossing from learning observations into psychological claims. Whether this constraint survives contact with real student data and institutional pressure is an empirical question.
 
Fifth, does meta-literacy training change how students use the architecture? Section \ref{sec:ethics} argued that co-authorship is a teachable skill, unevenly distributed, and that students who acquire it by accident enter AI-mediated learning at a growing advantage. f students taught to co-author show measurably different outcomes from those who are not, the architecture has a digital-divide problem and training becomes essential.
 
\bigskip
 
\noindent\textbf{Who designs the designer.} The designer's role, what the system treats as true about the student, has three occupants in the architecture proposed here. The educator sets the architecture: learning objectives, ethical boundaries, scaffolding. The student co-authors the rules: what the system treats as true, what is revised, what is removed. The AI executes within both. This is the minimum structure required to keep behavioural architecture from becoming a surveillance system. None of these three roles is currently occupied; naming that vacancy is the contribution of this paper.

\bibliographystyle{cas-model2-names}

\bibliography{cas-refs}

\end{document}